\documentclass[twocolumn,twoside,slac_two]{revtex4}
\usepackage{graphicx}
\usepackage{fancyhdr}
\newcommand{\be}[0]{\begin{equation}}
\newcommand{\ee}[0]{\end{equation}}
\newcommand{\ba}[0]{\begin{eqnarray}}
\newcommand{\ea}[0]{\end{eqnarray}}

\def\a{\alpha }

\begin{document}

\title{Simple model for QCD analysis of the proton helicity structure}

%

\author{Ali N. Khorramian}
\affiliation{Physics Department, Semnan University, Semnan, Iran}
\affiliation{Institute for Studies in Theoretical Physics and
Mathematics (IPM), P.O.Box 19395-5531, Tehran, Iran}
\author{S. Atashbar Tehrani}
\affiliation{Physics Department, Semnan University, Semnan, Iran}
\affiliation{Institute for Studies in Theoretical Physics and
Mathematics (IPM), P.O.Box 19395-5531, Tehran, Iran}

\begin{abstract}
In this paper we use the experimental data to obtain the polarized
parton distribution functions (PPDFs) in the LO and NLO
approximations. The analysis is based on the Jacobi polynomials
expansion of the polarized structure function (PSF). Our
calculations for polarized parton distribution functions based on
the Jacobi polynomials method are in good agreement with the other
theoretical models.

\end{abstract}

\maketitle

\thispagestyle{fancy}


\section{Introduction}
The theoretical and experimental status on the spin structure of
the nucleon  has been discussed in great detail in several recent
reviews (see, e.g.,
Refs.~\cite{Anselmino:1994gn,Lampe:1998eu,Hughes:1999wr,filippone-01}.
Deeply inelastic scattering provides a clean way to extract the
parton densities of nucleons. After the initial observation that
the nucleon spin is not formed by the quarks dominantly
\cite{SPUZ}, detailed measurements of the polarized structure
functions followed during the last 20~years. The central question
concerns now the parton distribution functions and their scale
evolution rather than just their first moment. Since the nucleon
spin receives also contributions from the angular momentum of the
quarks and gluons, these degrees of freedom have also to be
studied. During the recent years several comprehensive analysis of
the polarized deep inelastic scattering (DIS) data, based on
next-to-leading-order quantum chromodynamics have performed. In
these analysis the polarized parton density functions are either
written in terms of the well-known parameterizations of the
unpolarized PDFs or parameterized independently, and the unknown
parameters are determined by fitting the polarized DIS data.

Determination of  parton distributions in a nucleon in the
framework of quantum chromodynamics (QCD)  always involves some
model-dependent procedure. Instead of relying on mathematical
simplicity as a guide, we take a viewpoint in which  the physical
picture of the nucleon structure is emphasized. That is, we
consider the model for the nucleon which is compatible with the
description of the bound state problem in terms of three
constituent quarks. We adopt the view that these
 constituent quarks in the scattering problems should be regarded as the valence quark
clusters rather than point-like objects. They have been referred
to as $\it{valons}$. The idea of nucleon as a bound state of three
quarks was presented for the first time in Ref.
\cite{Altarelli:1973ff}. In the valon model, the proton consists
of two ``up'' and one ``down'' valons. These valons thus, carry
the quantum numbers of the respective valence quarks. Hwa and et
al. \cite{Hwa:2002mv} found evidence for the valons in the deep
inelastic neutrino scattering data, suggested their existence and
applied it to a variety of phenomena. In \cite{Arash:2003ci}
unpolarized PDFs and hadronic structure functions in the NLO
approximation were extracted. In Ref.~\cite{Khorramian:2004ih} the
polarized valon model is applied to determine the quark helicity
distributions and polarized proton structure functions in the NLO
approximation by using the Bernstein polynomial approach. The
extraction of the quark helicity distributions is one of the main
tasks of the semi-inclusive deep inelastic scattering (SIDIS)
experiments (HERMES \cite{Airapetian:2004zf}, COMPASS
\cite{compass}, SMC\cite{Adeva:1997qz}) with the polarized beam
and target. Recently in Ref.
\cite{Mirjalili:2006hf,Mirjalili:2007ep} the polarized valon model
was applied and analyzed the flavor-broken light sea quark
helicity distributions with the help of a Pauli-blocking ansatz.
The reported results of this paper are in good agreement with the
HERMES experimental data for the quark helicity distributions in
the nucleon for up, down, and strange quarks from semi-inclusive
deep-inelastic scattering \cite{Airapetian:2004zf}.

Since very recently experimental data are available from the
HERMES collaboration \cite{unknown:2006vy} for the spin structure
function $g_1$, therefore there is enough motivation to study and
utilize the spin structure and quark helicity distributions
extracted via the phenomenological model. We performed the first
recalculation for the spin structure function $g_1$ up to NLO
\cite{Atashbar Tehrani:2007be} using the Jacobi polynomials
expansion\cite{Khorramian:2007zza}. This paper is based on the
Ref.\cite{Atashbar Tehrani:2007be}.

The plan of the paper is to give a brief review of the theoretical
background of the QCD analysis in two loops. The method of the QCD
analysis of polarized structure function, based on Jacobi
polynomials are written down in Section 3. A description of the
procedure of the QCD fit of $g_1$ data and results are illustrated
in Section~4.
\section{The theoretical background}
Let us define the {\sc Mellin} moments for the polarized structure
function $g_1^p(x,Q^2)$:

\be g_1^p(N,Q^2)=\int_0^1 \;x^{N-1}g_1^p(x,Q^2)dx\;.
\label{momdefine} \ee The contribution to the structure function
$g_1(N,Q^2)$ up to NLO can be represented in terms of the
polarized parton densities
 and the coefficient functions $\Delta C_i^N$ in the
{\sc Mellin} -N space  by \cite{Lampe:1998eu}
\begin{eqnarray}
g_1^p(N,Q^2)&=&\frac{1}{2}\sum\limits_q
e^2_q\{(1+\frac{\alpha_s}{2\pi}\Delta C^N_q) [\Delta q(N,Q^2)+\nonumber \\
 &&\Delta\bar q(N,Q^2)] + \frac{\alpha_s}{2\pi}2\Delta
C^N_g\Delta g(N,Q^2)\}\;,\nonumber \\ \label{eq:momg1NLO}
\end{eqnarray}
in this equation the NLO running coupling constant is given by
\begin{eqnarray}
\label{eq:alfaNNLO}
A_s &=&\frac{1}{\beta _{0}\ln Q^{2}/\Lambda _{\overline{MS}%
}^{2}}-\frac{\beta _{1}\ln (\ln Q^{2}/\Lambda
_{\overline{MS}}^{2})}{\beta _{0}^{3}(\ln Q^{2}/\Lambda
_{\overline{MS}}^{2})^{2}}\;.\nonumber \\
\end{eqnarray}The symbol $A_s$
denotes the strong coupling constant normalized to
$A_s=\alpha_s/(4\pi)$. Notice that in the above the numerical
expressions for $\beta_0$, $\beta_1$ are
\begin{eqnarray}
\label{eq:beta}
\beta_0&=&11-0.6667f \;, \nonumber \\
\beta_1&=&102-12.6667f \;,\nonumber \\
\end{eqnarray}
where $f$ denotes the number of active flavors. In our calculation
, we choose $Q_0=1\;GeV^2$ as a fixed parameter and $\Lambda$ is
an unknown parameter which can be obtained  by fitting
to experimental data.\\

In Eq.~(\ref{eq:momg1NLO}), $\Delta
q(N,Q^2)=\Delta{{q_v(N,Q^2)}}+\Delta\bar q(N,Q^2)$, $\Delta\bar
q(N,Q^2)$ and $\Delta g (N,Q^2)$ are moments of the polarized
parton distributions in a proton. $\Delta C^N_q$, $\Delta C^N_g$
are also the $N$-th moments of spin-dependent Wilson coefficients
given in Ref.\cite{Atashbar Tehrani:2007be}.

According to improved polarized valon model framework,
determination of the moments of parton distributions in a proton
can be done strictly through the moments of the polarized valon
distributions.

The moments of PPDFs are denoted respectively by:
$\Delta{u_{v}}(N,Q^{2})$, $\Delta{d_{v}}(N,Q^{2})$, $\Delta{\Sigma
}(N,Q^{2})$ and $\Delta g(N,Q^{2})$. Therefore, the moments of the
polarized $u$, $d$, $\Sigma $ and $g$ density functions in a
proton are:
\begin{eqnarray}
\Delta{u_{v}}(N,Q^{2})=2\Delta M^{NS}(N,Q^{2})\times\Delta
M^{'}_{U/p}(N), \label{eq:momuv}
\end{eqnarray}
\begin{eqnarray}
\Delta{d_{v}}(N,Q^{2})=\Delta M^{NS}(N,Q^{2})\times\Delta M^{'}_
{D/p}(N), \label{eq:momdv}
\end{eqnarray}
\begin{equation}
\Delta{\Sigma }(N,Q^{2})=\Delta M^{S}(N,Q^{2})(2 \Delta
M^{''}_{U/p}(N)+\Delta M^{''}_{D/p}(N)), \label{eq:momsig}
\end{equation}
\begin{equation}
\Delta g(N,Q^{2})=\Delta M^{gq}(N,Q^{2})(2 \Delta
M^{'}_{U/p}(N)+\Delta M^{'}_{D/p}(N)).\nonumber \\
\end{equation}
In the above equations $M'_{j/p}(N)$ and  $M''_{j/p}(N)$ are the
moments of polarized valon distributions, which introduced in
Ref.\cite{Atashbar Tehrani:2007be}. It is obvious that the final
form for $g_1(N,Q^2)$ involves some unknown parameters. If the
parameters can be  obtained  then the computation of all moments
of the PPDFs and the polarized structure
function, $g_1(N,Q^2)$, are possible. \\\\\\\\

\begin{figure*}[t]
\centering \vspace{1 cm}
\includegraphics[width=95mm]{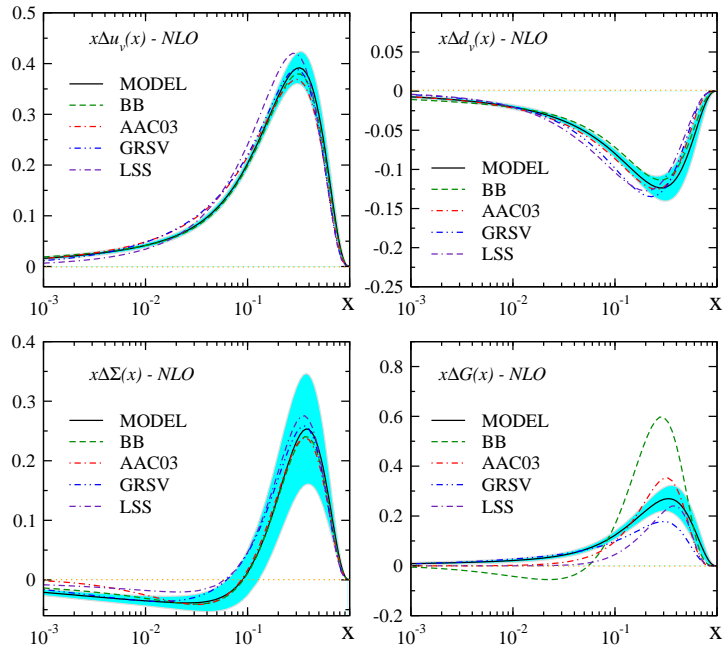}
\caption{NLO polarized parton distributions at the input scale
$Q_0^2$=1.0 GeV$^2$ compared to results obtained by BB model
(dashed line) (ISET=3)\cite{Blumlein:2002be}, AAC (dashed-dotted
line) (ISET=3)\cite{Goto:1999by}, GRSV (dashed-dotted dotted line)
(ISET=1)\cite{Gluck:2000dy} and LSS (dashed-dashed dotted line)
(ISET=1)\cite{Leader:2005ci}.} \label{JACpic2-f1}
\end{figure*}

\section{The method of the QCD analysis}
One of the simplest and fastest possibilities in the PSF
reconstruction from the QCD predictions for its Mellin moments is
Jacobi polynomials expansion. The Jacobi polynomials are
especially suited for this purpose since they allow one to factor
out an essential part of the $x$-dependence of the structure
function into the weight function \cite{parisi}. Thus, given the
Jacobi moments $a_{n}(Q^2)$, a structure function $f(x,Q^2)$ may
be reconstructed in a form of the series
\cite{Barker}-\cite{Barker:1980wu}
\begin{equation}
xf(x,Q^2)=x^{\beta}(1-x)^{\a} \sum_{n=0}^{N_{max}} a_{n}(Q^2)
\Theta_n ^{\a , \beta}(x), \label{e5}
\end{equation}
where $N_{max}$ is the number of polynomials and $\Theta_n ^{\a ,
\beta}(x)$ are the Jacobi polynomials of order $n$,
\begin{equation}
\Theta_{n} ^{\a , \beta}(x)= \sum_{j=0}^{n}c_{j}^{(n)}{(\a ,\beta
)}x^j , \label{e9}
\end{equation}
where $c_{j}^{(n)}{(\a ,\beta )}$ are the coefficients that
 expressed through $\Gamma$-functions and satisfy the orthogonality relation with the weight
$x^{\beta}(1-x)^{\a}$ as following

\begin{equation}
\int_{0}^{1}dx\;x^{\beta}(1-x)^{\a}\Theta_{k} ^{\a , \beta}(x)
\Theta_{l} ^{\a , \beta}(x)=\delta_{k,l}\ , \label{e8}
\end{equation}
For the moments, we note that the $Q^2$ dependence is entirely
contained in the Jacobi moments
\begin{eqnarray}
a_{n}(Q^2)&=&\int_{0}^{1}dx\;xf(x,Q^2)\Theta_{k} ^{\a ,
\beta}(x)\nonumber \\
&=&\sum_{j=0}^{n}c_{j}^{(n)}{(\a ,\beta )} f(j+2,Q^2) \;,
\label{e8nn}
\end{eqnarray}
obtained by inverting Eq.(~\ref{e5}), using Eqs.~(\ref{e9},
\ref{e8}) and also definition of moments,
$f(j,Q^2)=\int_{0}^{1}dx\;x^{j-2}xf(x,Q^2)$.
\\
Using  Eqs. (\ref{e5}-\ref{e8nn}) now, one can relate the PSF with
its Mellin moments \ba
xg_{1}^{N_{max}}(x,Q^2)&=&x^{\beta}(1-x)^{\a}
\sum_{n=0}^{N_{max}}\Theta_n ^{\a, \beta}(x)\times\nonumber\\
&&\sum_{j=0}^{n}c_{j}^{(n)}{(\a ,\beta )} g_{1}(j+2,Q^2),
\label{eg1Jacob} \nonumber \\\ea where $g_{1}(j+2,Q^2)$ are the
moments of polarized structure function. $N_{max}$, $\alpha$ and
$\beta$ have to be chosen so as to achieve the fastest convergence
of the series on the R.H.S. of Eq.~(\ref{eg1Jacob}) and to
reconstruct $xg_1$ with the required accuracy. In our analysis we
use $N_{max}=9$, $\alpha=3.0$ and $\beta=0.5$. The same method has
been applied to calculate the nonsinglet structure function $xF_3$
from their moments [24-31].

 Obviously the $Q^2$-dependence
of the polarized structure function is defined by the
$Q^2$-dependence of the moments.

\section{The procedure of the QCD fit and results}
For the QCD analysis presented in Ref.~\cite{Atashbar
Tehrani:2007be} the following data sets were used: the HERMES
proton data \cite{Airapetian:1998wi,unknown:2006vy}, the SMC
proton data \cite{Adeva:1998vv}, the E143 proton data
\cite{Abe:1998wq}, the EMC proton data
\cite{Ashman:1987hv,Ashman:1989ig}. In the fitting procedure,
using the CERN subroutine MINUIT \cite{James:1975dr}, we defined a
global ${\chi}^{2}$ for all the experimental data points and found
an acceptable fit with minimum ${\chi}^{2}/{\rm{d.o.f.}}=0.978$ in
the LO case and ${\chi}^{2}/{\rm{d.o.f.}}=0.933$ in the NLO case.
In Table. 2 of Ref.\cite{Atashbar Tehrani:2007be} we presented and
compared the results which are based on Bernstein and Jacobi
approaches.

In Figures~1 the parton distribution functions in  next-to-leading
order for all sets of parameterizations ~\cite{Goto:1999by,
Blumlein:2002be, Gluck:2000dy, Leader:2005ci} and their errors are
presented at the starting scale $Q_0^2$.

\begin{acknowledgments}
A.N.K is grateful to the Organizing Committee of the International
Conference on Hadron Physics TROIA'07, for financial support of
his participation in the conference. We would like to thank  Z.
Karamloo  for reading the manuscript of this paper. A.N.K. thanks
Semnan university for partial financial support of this project.

\end{acknowledgments}

\bigskip 

\end{document}